# Simultaneous Multi-Anomaly Detection Using Quantum Kernel-Based Machine Learning

Takao Tomono[1], Kazuya Tsujimura[2]

[1] Research and Development Directorate, Japan Aerospace Exploration Agency (JAXA), Japan

[2] Digital Innovation Division, TOPPAN Holdings Inc., Japan

## Abstract

Maintaining production equipment is essential in manufacturing, yet the computational cost of machine-learning-based monitoring increases with the number of machines. Autoregressive (AR) model coefficients combined with one-class support vector machines (OCSVMs) are widely used for anomaly detection in vibration and acoustic signals. This study extends our previous work from single-anomaly detection to simultaneous multi-anomaly scenarios and evaluates quantum kernel methods in both settings. Five miniature semiconductor manufacturing devices were operated simultaneously, and ten anomaly types were intentionally induced. Acoustic signals were captured using a single microphone, segmented into analysis windows, and modeled with sixth-order AR processes. The resulting AR coefficients were classified using OCSVMs with three quantum kernels and compared with a classical radial basis function (RBF) kernel OCSVM and an Isolation Forest. For single-anomaly detection, the best quantum kernel achieved an F2 score of 0.994 and an AUROC of 0.976. For simultaneous detection of two anomalies across five anomaly pairs, performance decreased but remained superior to classical baselines, achieving an F2 score of 0.872 and an AUROC of 0.901. A modified Mahalanobis-distance-based feature space enabled discrimination among anomaly types, while simultaneous anomalies exhibited nonlinear acoustic interactions. These results demonstrate the potential of quantum kernel methods for practical multi-anomaly detection in industrial equipment monitoring.

## 1. Introduction

Predictive maintenance of production equipment underpins product quality, productivity, and safety in manufacturing. Continuous monitoring that detects anomalies before failures occur can prevent defective products from being shipped and avoid unplanned downtime. As the number of monitored machines grows, however, the computational cost of conventional machine learning approaches increases rapidly. Sensors and models must be prepared for each machine and each failure mode, and when several anomalies can occur at the same time, the number of combinations to be considered explodes: choosing two of ten anomaly types already gives 45 combinations, and choosing three gives 120.

In vibration and acoustic analysis, AR model coefficients combined with kernel methods such as the one-class SVM (OCSVM) are widely used for anomaly detection [1]–[3]. AR coefficients summarize the spectral structure of a signal with only a few parameters, and the OCSVM can be trained on normal data alone, which suits manufacturing sites where anomalous data are scarce. Separating several anomalies from mixed sound, however, requires a kernel that can represent complex combinations of features.

Quantum kernel methods embed classical data into a quantum Hilbert space through a quantum feature map and use the resulting kernel in a classical learner [4]–[6]. We have applied quantum kernels to anomaly detection in visual and shipping inspection [7]–[9]. For non-contact acoustic monitoring, we showed that a single microphone and a quantum kernel can classify two anomalies occurring in two machines [10], and that ten individual anomaly types in five machines can all be classified and identified [11]. Both studies, however, assumed that anomalies occur one at a time; whether quantum kernels remain effective when several anomalies occur simultaneously has not been examined.

In this paper, we extend our single-anomaly study to simultaneous multi-anomaly scenarios and investigate the advantages and limitations of quantum kernel methods in both settings. Our contributions are as follows:

(1) Using the same pipeline (AR features and quantum-kernel OCSVM), we evaluate ten individual anomaly types and five pairs of simultaneous anomalies, and compare three quantum kernels with an RBF-kernel OCSVM and an Isolation Forest under the same two-stage decision rule. The best quantum kernel (QK25) outperforms both classical baselines in threshold-dependent ($F_\beta$) and threshold-independent (AUROC) metrics.

(2) We show that a modified Mahalanobis metric space distinguishes anomaly types by their positions for both individual and simultaneous anomalies.

(3) By comparing observed centroid shifts with the prediction of linear superposition, we show that simultaneous anomalies do not lie at the superposition of individual anomaly positions, which suggests nonlinear acoustic interactions in the measurement environment.

Section 2 reviews related work, Section 3 describes the method, and Section 4 presents the experimental setup. Section 5 reports the results, Section 6 discusses them, and Section 7 concludes the paper.

## 2. Related Work

### 2.1 AR features and one-class SVMs for anomaly detection

An AR model expresses a time series as a linear combination of its past values, and its coefficients summarize the spectral structure of the signal with a few parameters. Cheng et al. [1] used AR coefficients as features for SVM-based gear fault diagnosis, and Wang and Makis [2] applied a statistical test to AR residuals to detect gear-shaft faults. Compared with other common acoustic features (Table 1), AR models are computationally light, have a single hyperparameter (the order p), and work with small data. Their inability to model nonlinearity directly can be compensated by kernel methods.

**Table 1.** Comparison of acoustic feature-extraction methods (◎ excellent, ○ good, △ fair, × poor).

| Item | AR model | FFT | STFT | Wavelet | MFCC | Autoencoder |
|---|---|---|---|---|---|---|
| Computation time | ◎ < 1 ms | ◎ < 1 ms | ○ a few ms | ○ ms to tens of ms | ○ a few ms | △ s to min |
| Interpretability | ◎ high | ◎ high | ◎ high | ○ medium | ○ medium | × low |
| Data required | ◎ small | ◎ small | ◎ small | ◎ small | ◎ small | × large |
| Tuning | ◎ order p only | ◎ almost none | ○ window | △ mother wavelet | △ frame, coefficients | × layers, dims, rate |
| Nonlinearity | × linear | × harmonics only | × limited | △ local | △ partial | △ applicable |
| Kernel compatible | ○ yes | no | no | ○ yes | ○ yes | △ redundant |

Because failures are rare in well-maintained equipment, sufficient anomalous data are difficult to collect. The OCSVM [3] learns a decision boundary from normal data only and flags samples outside it as anomalous, so it can also respond to anomaly combinations that were never seen during training. The Isolation Forest [12] is a widely used non-kernel alternative that isolates anomalies by random axis-parallel partitioning; we use it as a classical baseline.

### 2.2 Quantum kernel methods

Havlíček et al. [5] proposed supervised learning with quantum-enhanced feature spaces, and Schuld and Killoran [4] formulated quantum machine learning as kernel methods in feature Hilbert spaces. Huang et al. [6] showed that, with enough data, classical models can rival quantum models, so any quantum advantage depends on the data and the geometry of the kernel. Quantum kernels must therefore be evaluated against classical kernels on the target data.

### 2.3 Mahalanobis distance

The Mahalanobis distance measures how far a point lies from the center of a distribution while accounting for the covariance between features [13]; it is widely used for multivariate outlier detection and underlies the Mahalanobis–Taguchi system in quality engineering [14]. We use it not only to flag outliers but also to identify anomaly classes.

### 2.4 Our previous work

In [10], we identified the source machine of an anomaly from the quadrant of a two-dimensional feature plane in which it appeared, an approach that does not scale to many anomalies and features. In [11], a six-qubit quantum-kernel OCSVM classified ten individual anomaly types in five machines with $F_{0.5} = F_1 = F_2 = 1.0$ per anomaly class, and a modified Mahalanobis space identified them. The present study extends this single-anomaly setting to simultaneous anomalies and adds classical baselines.

## 3. Method

### 3.1 Overview

Fig. 1 shows the pipeline. Sound from all machines is recorded with a single microphone, segmented, and converted to AR coefficients. A kernel OCSVM trained on normal data only classifies each sub-segment, and a two-stage rule produces the sample-level decision. Samples judged anomalous are then projected onto a modified Mahalanobis space for identification.

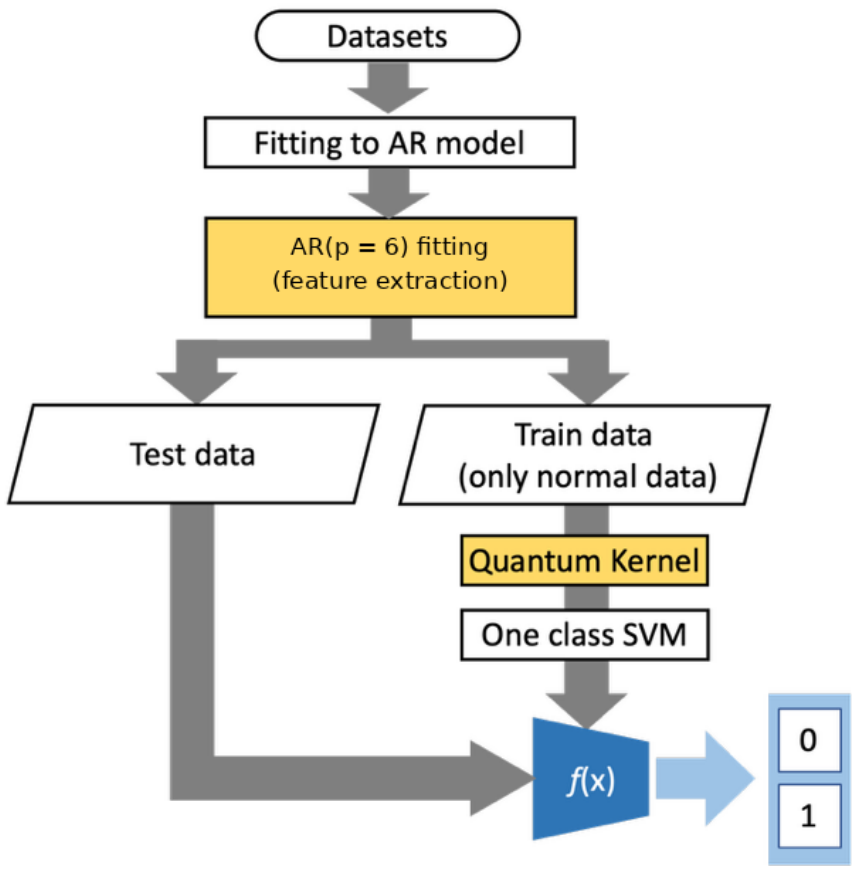

**Fig. 1.** Overview of the proposed pipeline.

### 3.2 Preprocessing and segmentation

Each condition consists of one 5-min recording (channel 2 of the microphone, sampled at 44.2 kHz), which is divided into 60 non-overlapping 5-s samples. Samples containing impulsive noise are excluded using an interquartile-range (IQR) rule: lower and upper quantiles ($Q_1$, $Q_3$) of the waveform amplitudes in the sample are computed, points outside [$Q_1 - 1.5$ IQR, $Q_3 + 1.5$ IQR] are counted as outliers, and the sample is excluded when the outlier ratio is at least $10^{-5}$. Instead of the usual quartiles (25th and 75th percentiles), we use the wider 15th and 85th percentiles.

The quantile range was widened because the amplitude of the machine sound varied greatly. With the usual quartiles, points beyond about $\pm 2.7\sigma$ from the mean would be counted as outliers under a Gaussian assumption, so ordinary amplitude fluctuations in machine sound would be flagged and many samples excluded. With the 15th and 85th percentiles, the outlier range starts at about $\pm 4.1\sigma$, so only samples with obvious impulsive noise are excluded. Avoiding data loss was important because the test data are limited to about 600 samples per condition, and excluding transient sound changes caused by the anomalies themselves would remove the very information to be detected and bias the evaluation. The rule uses no labels and is applied identically to training and test data. Note that these quantiles are parameters of the exclusion rule and are distinct from the confidence intervals of the evaluation metrics (Section 4.4).

The rule excluded 6 of 60 normal test samples, 1 of 600 single-anomaly samples, and 2 of 300 simultaneous-anomaly samples. Each remaining 5-s sample is divided into 20 sub-segments of 0.25 s; the horizontal axis of Fig. 5 is the number of these sub-segments judged anomalous.

### 3.3 AR feature extraction

Each sub-segment $y_t$ is modeled by an AR(p) process without a constant term,

$$y_t = \sum_{i=1}^{p} \varphi_i y_{t-i} + \varepsilon_t \tag{1}$$

where $\varphi_i$ are the AR coefficients and $\varepsilon_t$ is white noise. The coefficients are estimated by ordinary least squares using statsmodels [15]. Following a preliminary comparison of the Akaike and Bayesian information criteria [16] over p = 3–7 [11], we use p = 6, except in Section 5.2, where p is varied. Each 5-s sample is thus represented by a 20 × 6 matrix of AR coefficients. Features are standardized with a scaler fitted to the training data.

### 3.4 Kernel one-class SVM

The OCSVM finds a hyperplane that separates normal data from the origin in the kernel-induced feature space [3]; we use the scikit-learn implementation [17] with $\nu = 0.1$. The classical baseline uses the RBF kernel of Eq. (2) with $\gamma = 0.1$.

$$k_{RBF}(x, x') = exp(-\gamma \|x - x'\|^2) \tag{2}$$

The quantum kernel embeds a feature vector x into a quantum state with a feature map U(x) and is defined as the fidelity between two states [4], [5]:

$$k_Q(x, x') = |\langle 0^{\otimes n}| U^{\dagger}(x') U(x) |0^{\otimes n}\rangle|^2 \tag{3}$$

where n = 6 is the number of qubits, each encoding one AR coefficient. Kernel matrices were computed from state vectors with the Qiskit statevector simulator (version 0.42.0) and passed to the OCSVM as precomputed kernels.

### 3.5 Quantum feature maps

We use three feature maps (Fig. 2, Table 2). QK23 has one-directional staircase entanglement with all Rz rotations placed after a barrier, separating the roles of Ry (strengthening entanglement) and Rz (phase adjustment). QK24 adds reverse CNOTs and feedback to the first qubit, giving bidirectional entanglement. QK25 keeps the QK24 structure and applies the same Ry rotation twice in succession on each qubit. Because consecutive rotations about the same axis add, $Ry(\theta)Ry(\theta) = Ry(2\theta)$, so that $U_{25}(x) = U_{24}(2x)$ and $k_{25}(x, x') = k_{24}(2x, 2x')$. QK25 is therefore not a new circuit structure but QK24 with the input features scaled by a factor of two, which corresponds to changing the bandwidth of the quantum kernel [18], [19]. After the two rotations are merged, QK24 and QK25 generate the same family of quantum states and have the same circuit depth. Because the features are standardized, most values lie within about ±3; doubling them yields rotation angles of up to about ±6 rad (nearly one full turn), so the periodicity of the rotations can map different inputs to similar states.

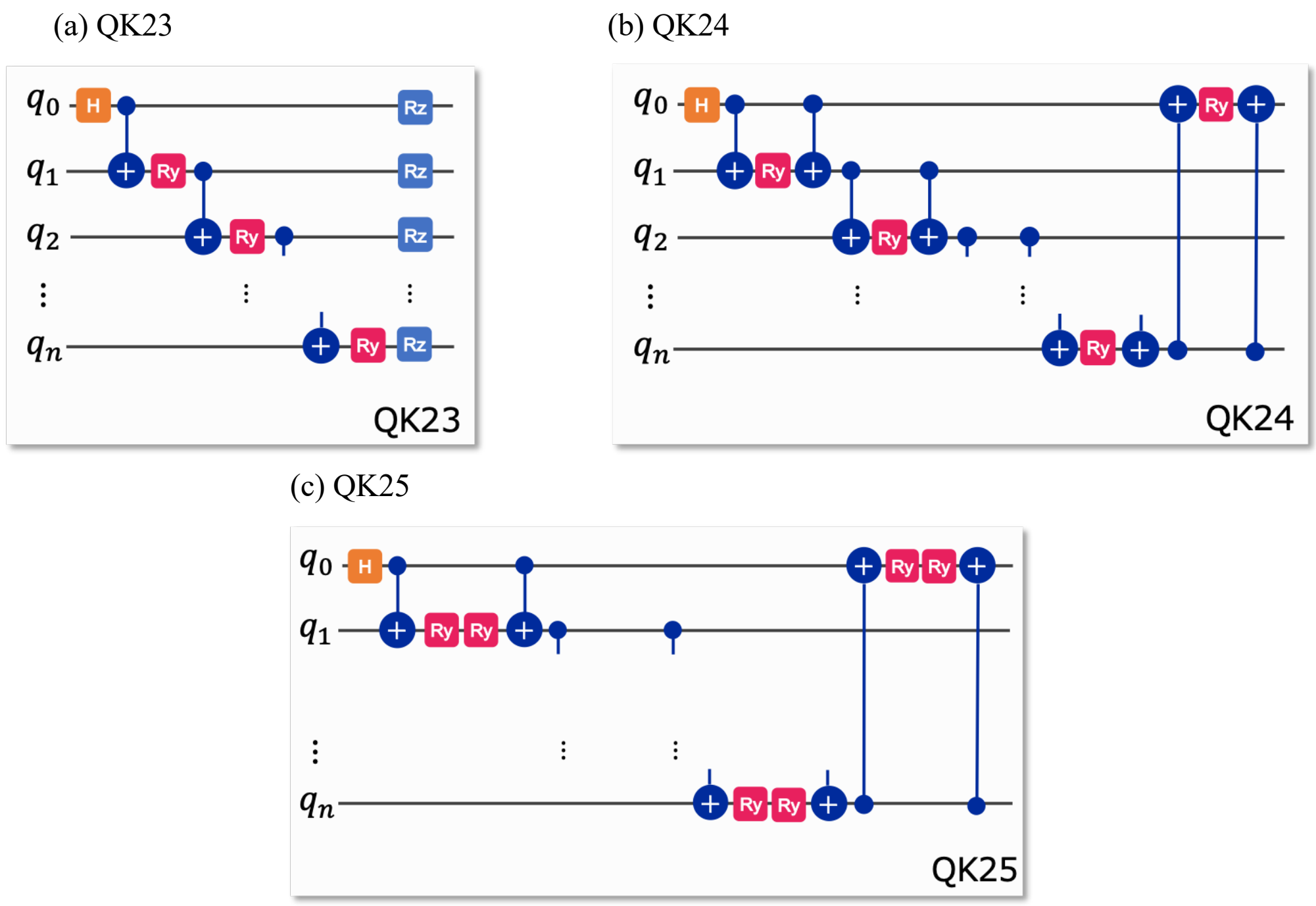

**Fig. 2.** Quantum feature maps (H: Hadamard gate; Ry, Rz: rotation gates; ● and ⊕: CNOT gate).

**Table 2.** Structure of the quantum feature maps.

| Kernel | Entanglement | Rotation gates | Depth |
|---|---|---|---|
| QK23 | One-directional staircase ($q_0 \rightarrow q_1 \rightarrow \ldots \rightarrow q_n$) | Ry after each CNOT; Rz grouped after a barrier | Shallow |
| QK24 | Bidirectional (reverse CNOTs and feedback to $q_0$) | Ry distributed; Ry on $q_0$ last | Deep |
| QK25 | Bidirectional (as QK24) | Same Ry applied twice (≡ $Ry(2\theta)$) | Same as QK24 |

## 3.6 Two-stage decision rule

Detection is performed in two stages. First, each of the 20 sub-segments of a sample is classified as normal or anomalous by the OCSVM (sign of the decision function). Second, a sample is declared anomalous when at least 15 of its 20 sub-segments (75%) are anomalous. The number of anomalous sub-segments (0–20) also serves as a sample-level anomaly score, and the mean decision-function value over the 20 sub-segments serves as a second, fully threshold-independent score (Section 4.4).

### 3.7 Identification in a modified Mahalanobis space

For samples judged anomalous, we take the fourth and fifth standardized AR coefficients ($\varphi_4$, $\varphi_5$) as axes and compute the centroid $\mu_k$ of each anomaly class k and the centroid $\mu_0$ of the normal data. The centroid shift, divided by the spread $\sigma_k$ of class k, gives the coordinates

$$x^m = \frac{|\mu_{k,4} - \mu_{0,4}|}{\sigma_k}, \quad y^m = \frac{|\mu_{k,5} - \mu_{0,5}|}{\sigma_k} \tag{4}$$

where $\sigma_k$ is the mean of the standard deviations of class k along the two axes. The 4th and 5th features correspond to $\varphi_4$ and $\varphi_5$. The coordinates ($x^m$, $y^m$) are plotted (Fig. 3). The origin corresponds to the normal centroid; classes far from it are easier to distinguish from normal operation, and classes at distinct positions can be identified by their positions.

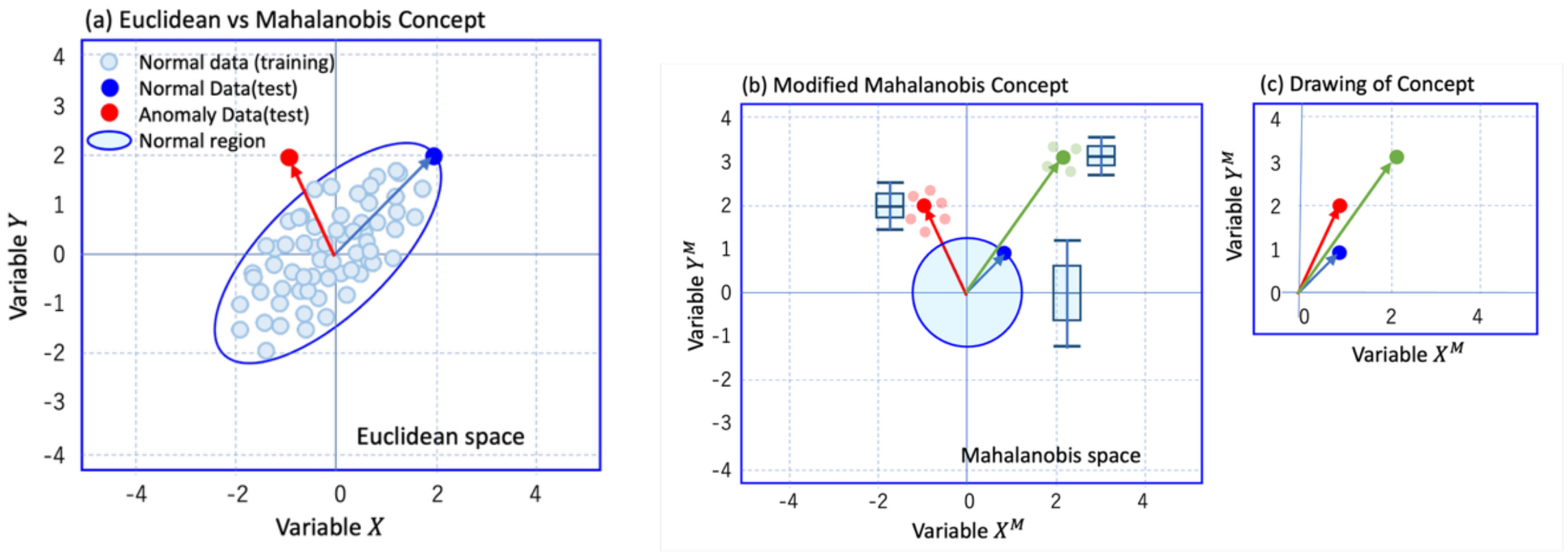

**Fig. 3.** Concept of the modified Mahalanobis space. (a) Euclidean versus Mahalanobis distance; (b) modified Mahalanobis space; (c) its representation in the first quadrant.

## 4. Experimental Setup

### 4.1 Dataset

Five miniature pieces of semiconductor factory equipment were installed in a simulated factory: machines S (S1, S2), C (C1, C2), and P (P1) (Fig. 4). Sound was recorded with a single microphone while all five machines operated simultaneously. Details of the machines, their positions, and their acoustic characteristics will be reported separately.

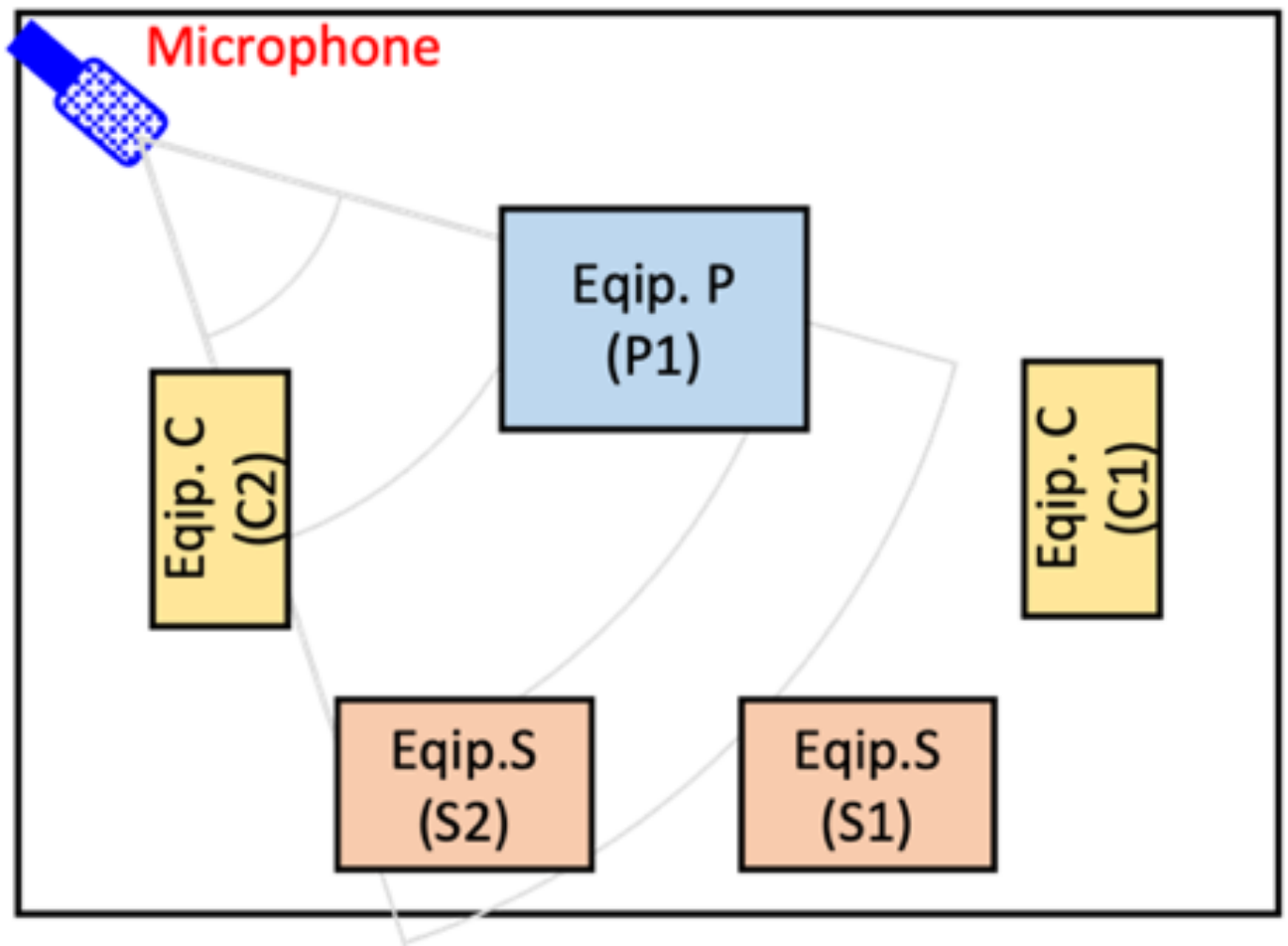

**Fig. 4.** Layout of the machines and microphone in the simulated factory (reproduced from [11]).

Ten anomaly types were intentionally induced in different components: three each in S1 (S1-1, S1-2, S1-3) and S2 (S2-1, S2-2, S2-3), and two each in C1 (C1-1, C1-2) and P1 (P1-1, P1-2). C2 was always operated normally. The mean signal-to-noise ratio computed directly from the recorded waveforms was −31.2 dB for normal operation and −32.3 to −32.9 dB for the ten anomalies.

### 4.2 Experimental conditions

Three conditions were recorded:

(1) **Normal (Exp. 1):** all five machines operate normally; used for training (one 5-min recording, 60 samples).

(2) **Single anomaly (Exp. 2):** one of the ten anomaly types occurs; one 5-min recording for each anomaly type and for normal operation; test set of 653 samples (54 normal; 599 anomalous from 10 types × 60 samples).

(3) **Simultaneous anomalies (Exp. 3):** two anomaly types occur simultaneously in five pairs (S1-1 + C1-1, P1-1 + C1-1, S1-1 + P1-1, S2-1 + P1-1, S1-3 + S2-1); one 5-min recording per pair; test set of 352 samples (54 normal; 298 anomalous from 5 pairs × 60 samples).

All models were trained on Exp. 1 data only and tested on the normal and anomalous samples of each condition.

### 4.3 Classical baselines

The RBF-kernel OCSVM uses the same features, scaler, ν, and two-stage rule as the quantum kernels. The Isolation Forest [12] uses 300 trees, all training samples per tree, and all six features. Its sub-segment threshold is the 90th percentile of out-of-fold anomaly scores of the normal training data (5-fold), which matches the false-alarm level of the OCSVM (ν = 0.1), and the same 15-of-20 rule is applied. As a sensitivity check, we also evaluated the 95th percentile.

### 4.4 Evaluation metrics

Treating anomalies as positive, precision and recall are given by Eq. (5) and the $F_\beta$ score by Eq. (6), where TP, FP, and FN are the numbers of true positives, false positives, and false negatives.

$$Precision = \frac{TP}{TP+FP}, \quad Recall = \frac{TP}{TP+FN} \tag{5}$$

$$F_\beta = \frac{(1+\beta^2)\cdot Precision\cdot Recall}{\beta^2\cdot Precision+Recall} \tag{6}$$

$F_{0.5}$ emphasizes avoiding false alarms, $F_2$ emphasizes avoiding missed anomalies, and $F_1$ weighs both equally; reporting all three reveals whether a model is biased toward the normal or the anomalous side. Because preventing defective products from reaching customers has priority on production lines, we focus on $F_2$. We also report specificity (recall of the normal class) and threshold-independent AUROC and AUPRC computed from (A1) the number of anomalous sub-segments and (A2) the mean decision-function value over the 20 sub-segments. 95% confidence intervals of $F_2$ were obtained by bootstrap resampling of the confusion-matrix counts from a multinomial distribution (10,000 resamples; 2.5th–97.5th percentiles).

## 5. Results

### 5.1 Separation of normal and anomalous samples

Fig. 5 shows, for Exp. 2, the number of sub-segments judged anomalous in each sample. The horizontal axis is the number of the 20 sub-segments of each sample that are judged anomalous. With quantum kernels, normal samples concentrate at small counts and anomalous samples at large counts, so the threshold of 15 (75% of the sub-segments) separates them. Separation improves in the order QK23 < QK24 < QK25. With the RBF kernel, anomalous samples spread over the entire range and overlap with normal samples.

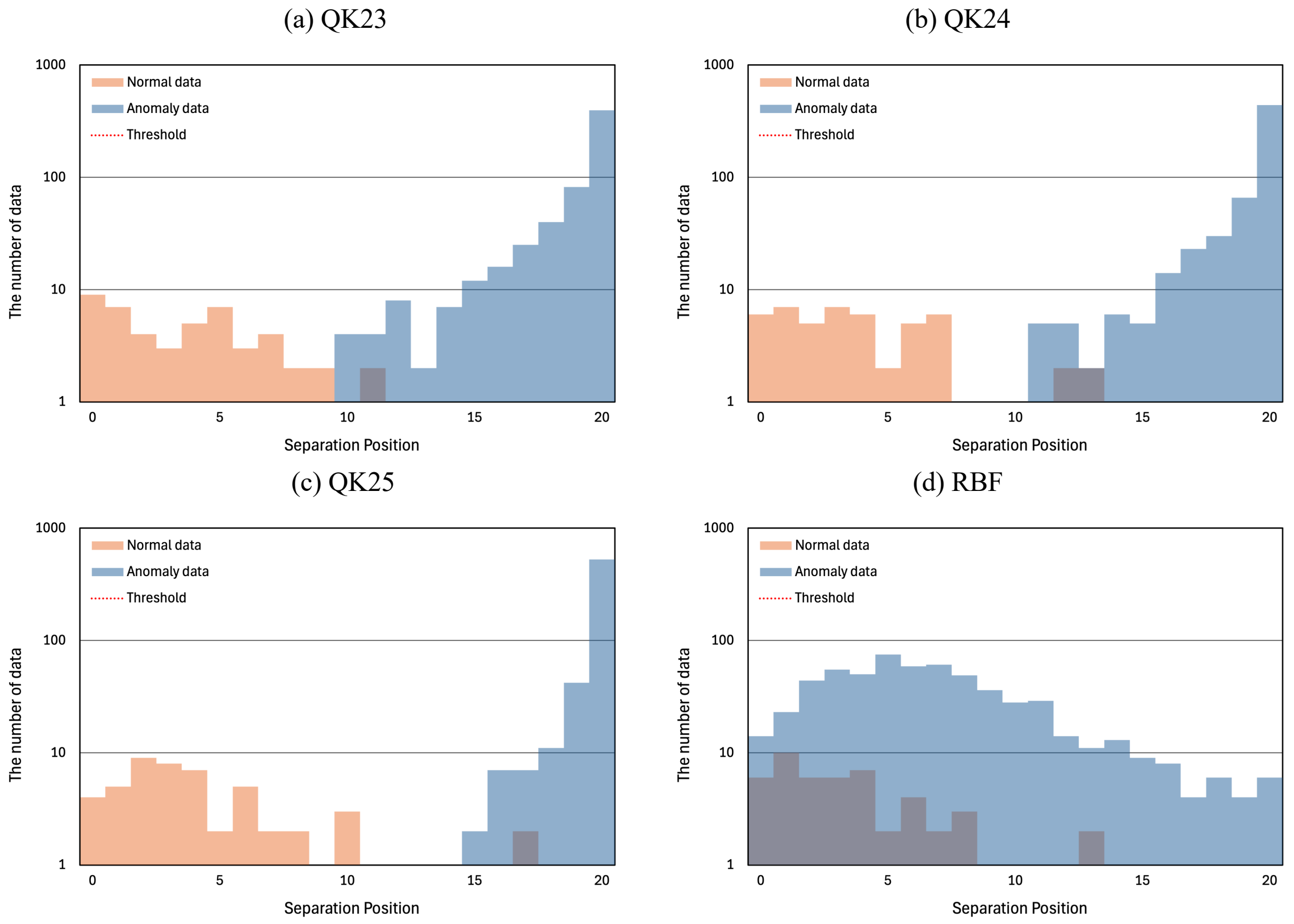

**Fig. 5.** Distribution of the number of anomalous sub-segments per sample (horizontal axis, 0–20) in Exp. 2. Light red: normal test data; light blue: anomalous test data. The vertical axis is the number of samples (logarithmic). The decision threshold is 15 (75% of the 20 sub-segments).

## 5.2 Effect of the AR order

Fig. 6 shows the $F_1$ score for p = 3–7. The quantum kernels achieve high $F_1$ at p = 5–6, whereas the RBF kernel is nearly constant for p = 3–5 and declines for p ≥ 6. Overall, the quantum kernels outperform the RBF kernel, although QK25 falls below RBF at p = 3. We use p = 6 hereafter.

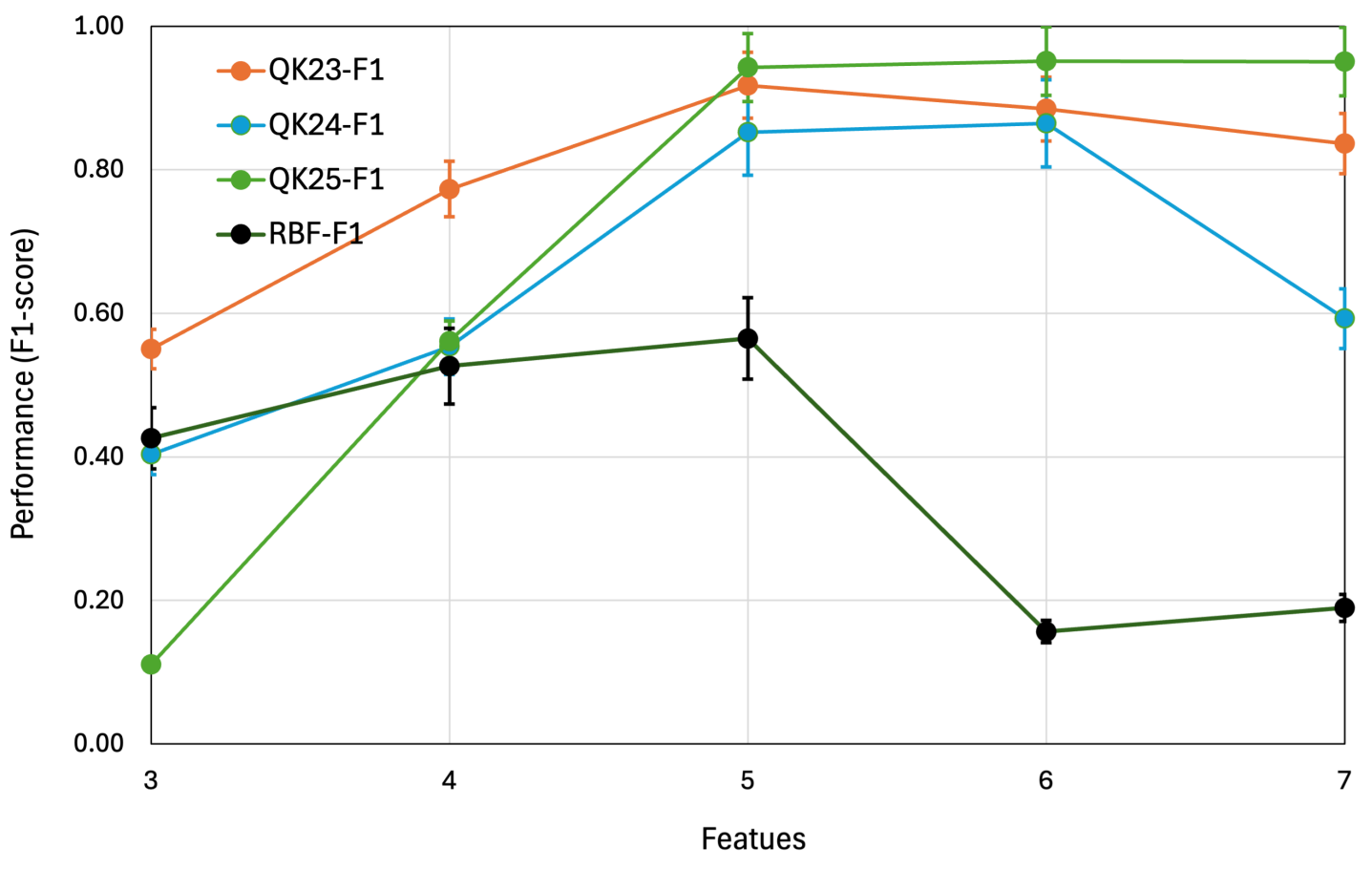

**Fig. 6.** $F_1$ score versus AR order p in Exp. 2.

### 5.3 Detection of individual and simultaneous anomalies

Table 3 summarizes the sample-level performance at the operating threshold, including 95% confidence intervals (CIs) for $F_2$, and Table 4 gives the threshold-independent metrics. In Exp. 2, QK25 achieved $F_2 = 0.994$ (95% CI 0.989–0.998) with precision 0.992 and recall 0.995, followed by QK24 ($F_2 = 0.976$, 0.964–0.986). QK23 kept high precision but missed many anomalies (recall 0.559, $F_2 = 0.612$). Both classical baselines were conservative: they rarely raised false alarms (specificity 0.963) but detected only about 36% of anomalous samples (RBF: $F_2 = 0.410$, 0.369–0.450; Isolation Forest: $F_2 = 0.415$, 0.374–0.456). The CIs of QK25 and QK24 overlap neither with each other nor with those of the baselines. Per anomaly class (Fig. 7), QK25 attained $F_{0.5} = F_1 = F_2 = 1.0$ for all ten anomalies [11], whereas the RBF kernel reached $F_2 \approx 0.8$ only for S1-2 and remained at or below 0.4 for most anomalies.

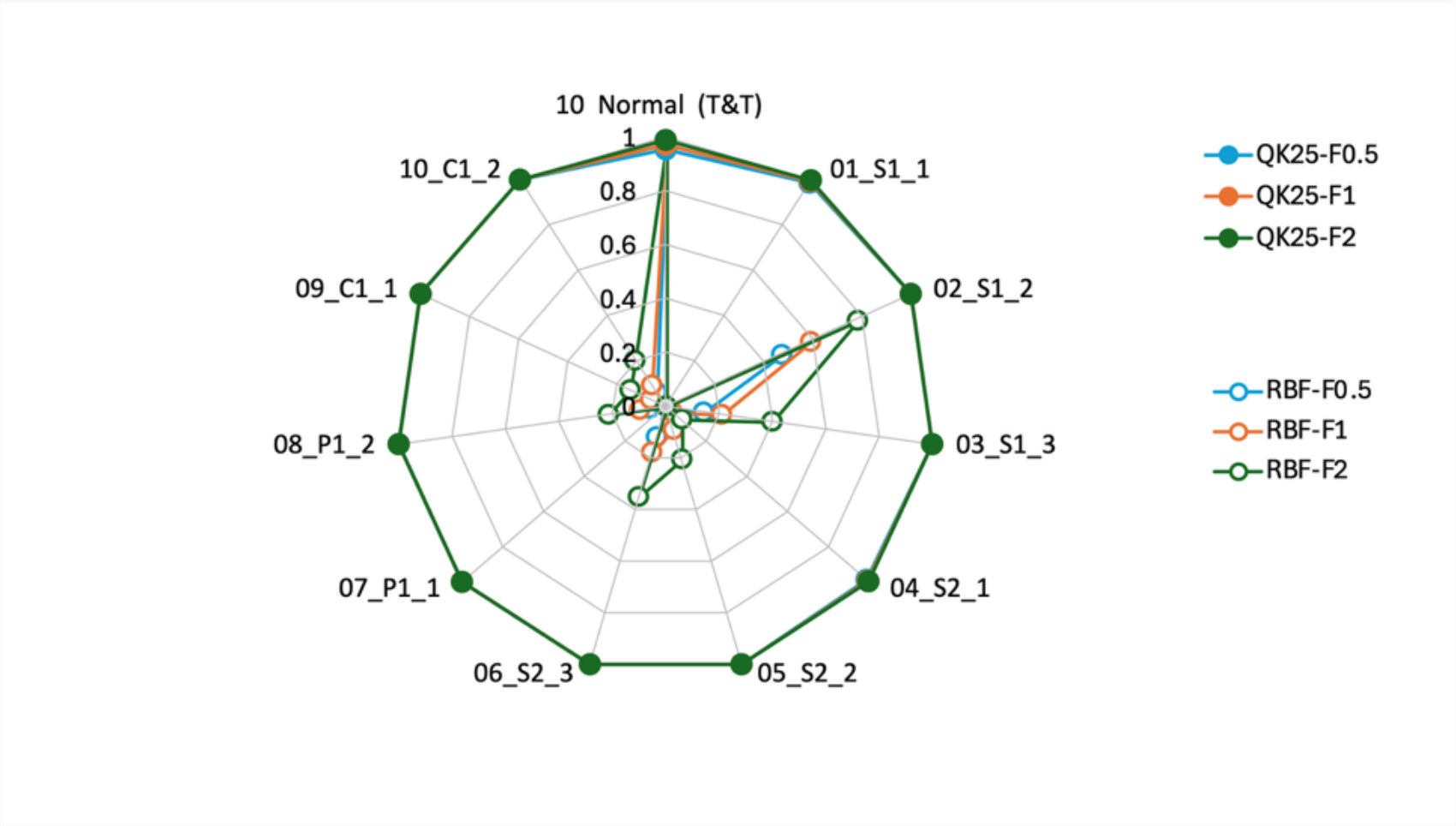

**Fig. 7.** $F_{0.5}$, $F_1$, and $F_2$ scores for each of the ten individual anomalies (filled: QK25; open: RBF) [11].

**Table 3.** Sample-level detection performance at the operating threshold (anomalous = positive).

| Method | Precision | Recall | Specificity | $F_{0.5}$ | $F_1$ | $F_2$ | $F_2$ 95% CI |
|---|---|---|---|---|---|---|---|
| **Exp. 2: single anomaly** (n = 653) | | | | | | | |
| QK25 | 0.992 | **0.995** | 0.907 | **0.992** | **0.993** | **0.994** | [0.989, 0.998] |
| QK24 | **0.993** | 0.972 | 0.926 | 0.989 | 0.982 | 0.976 | [0.964, 0.986] |
| QK23 | 0.988 | 0.559 | 0.926 | 0.857 | 0.714 | 0.612 | [0.574, 0.650] |
| RBF | 0.991 | 0.357 | **0.963** | 0.731 | 0.525 | 0.410 | [0.369, 0.450] |
| Isolation Forest (90th) | 0.991 | 0.362 | **0.963** | 0.736 | 0.531 | 0.415 | [0.374, 0.456] |
| **Exp. 3: two simultaneous anomalies** (n = 352) | | | | | | | |
| QK25 | 0.981 | **0.849** | 0.907 | **0.951** | **0.910** | **0.872** | [0.836, 0.905] |
| QK24 | 0.982 | 0.711 | 0.926 | 0.912 | 0.825 | 0.753 | [0.705, 0.798] |
| QK23 | 0.980 | 0.668 | 0.923 | 0.896 | 0.794 | 0.713 | [0.663, 0.761] |
| RBF | 0.984 | 0.416 | **0.963** | 0.773 | 0.585 | 0.470 | [0.413, 0.530] |
| Isolation Forest (90th) | **0.985** | 0.433 | **0.963** | 0.785 | 0.601 | 0.488 | [0.431, 0.544] |

**Table 4.** Threshold-independent performance (anomalous = positive). A1: number of anomalous sub-segments; A2: mean decision-function value.

| Method | AUROC (A1) | AUROC (A2) | AUPRC (A1) | AUPRC (A2) |
|---|---|---|---|---|
| **Exp. 2** (AUPRC chance level 0.917) | | | | |
| QK25 | **0.989** | **0.976** | **0.998** | **0.997** |
| QK24 | 0.980 | 0.920 | 0.997 | 0.989 |
| QK23 | 0.901 | 0.786 | 0.985 | 0.972 |
| RBF | 0.785 | 0.770 | 0.971 | 0.972 |
| Isolation Forest (90th) | 0.843 | 0.864 | 0.981 | 0.985 |
| **Exp. 3** (AUPRC chance level 0.847) | | | | |
| QK25 | **0.942** | **0.901** | **0.986** | **0.977** |
| QK24 | 0.880 | 0.823 | 0.972 | 0.960 |
| QK23 | 0.904 | 0.913 | 0.976 | 0.983 |
| RBF | 0.715 | 0.710 | 0.932 | 0.937 |
| Isolation Forest (90th) | 0.775 | 0.799 | 0.951 | 0.958 |

In Exp. 3, performance decreased for every method, but QK25 remained the best ($F_2$ = 0.872, 95% CI 0.836–0.905; AUROC = 0.901), followed by QK24 ($F_2$ = 0.753, 0.705–0.798). The RBF kernel and the Isolation Forest detected 42% and 43% of anomalous samples, respectively ($F_2$ = 0.470, 0.413–0.530; and 0.488, 0.431–0.544). In the threshold-independent metrics, QK25 and QK24 also exceeded both baselines in both experiments (Table 4), although the Isolation Forest attained a higher AUROC (A2) than QK23 in Exp. 2 (0.864 vs. 0.786). With the stricter 95th-percentile threshold, the Isolation Forest detected fewer anomalies ($F_2$ = 0.208 in Exp. 2 and 0.355 in Exp. 3), so the comparison does not depend on this choice. Fig. 8 and Table 5 show $F_2$ per pair: QK25 achieved $F_2 \geq 0.8$ for all pairs, whereas the RBF kernel exceeded 0.1 only for P1-1 + C1-1.

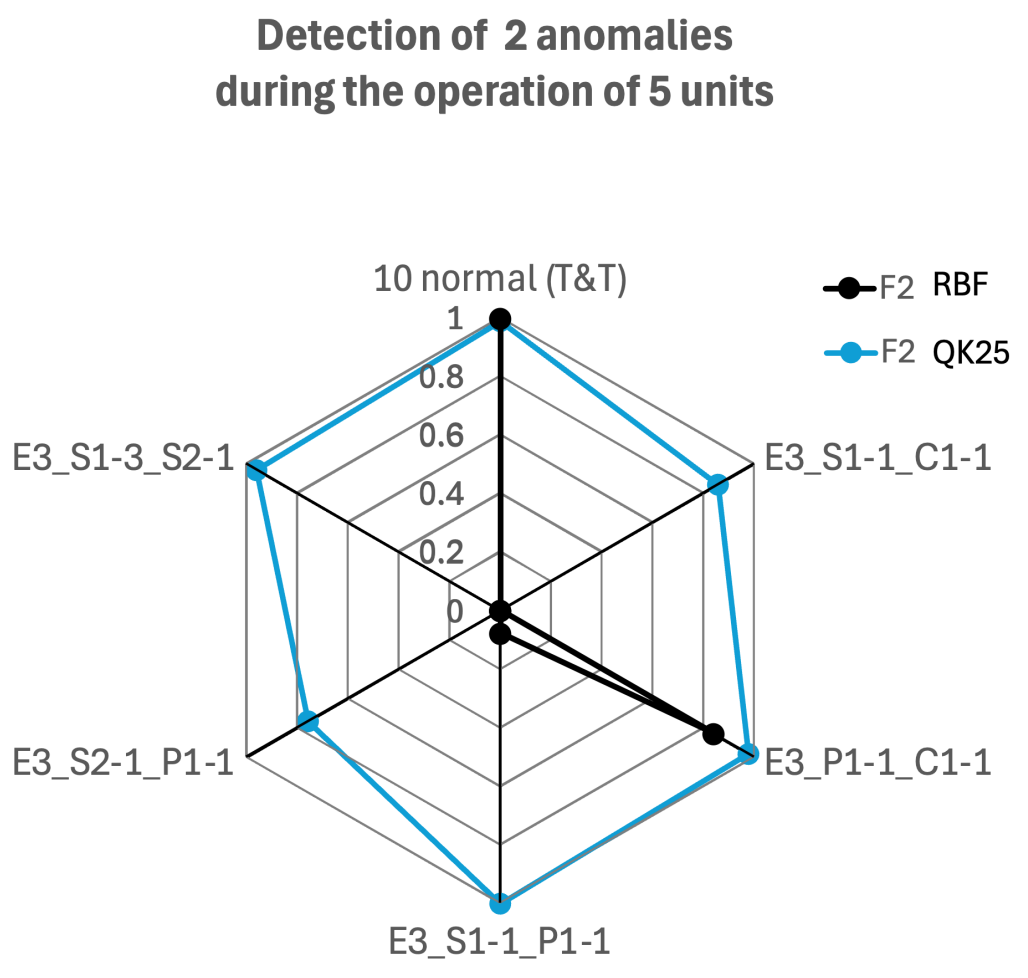

**Fig. 8.** $F_2$ scores for each pair of simultaneous anomalies (QK25 and RBF).

**Table 5.** $F_2$ per pair of simultaneous anomalies.

| Pair | QK25 | RBF |
|---|---|---|
| P1-1 + C1-1 | ≈ 1.0 | 0.9 |
| S1-1 + P1-1 | ≈ 1.0 | < 0.1 |
| S1-3 + S2-1 | ≈ 1.0 | < 0.1 |
| S1-1 + C1-1 | 0.8 | < 0.1 |
| S2-1 + P1-1 | 0.8 | < 0.1 |

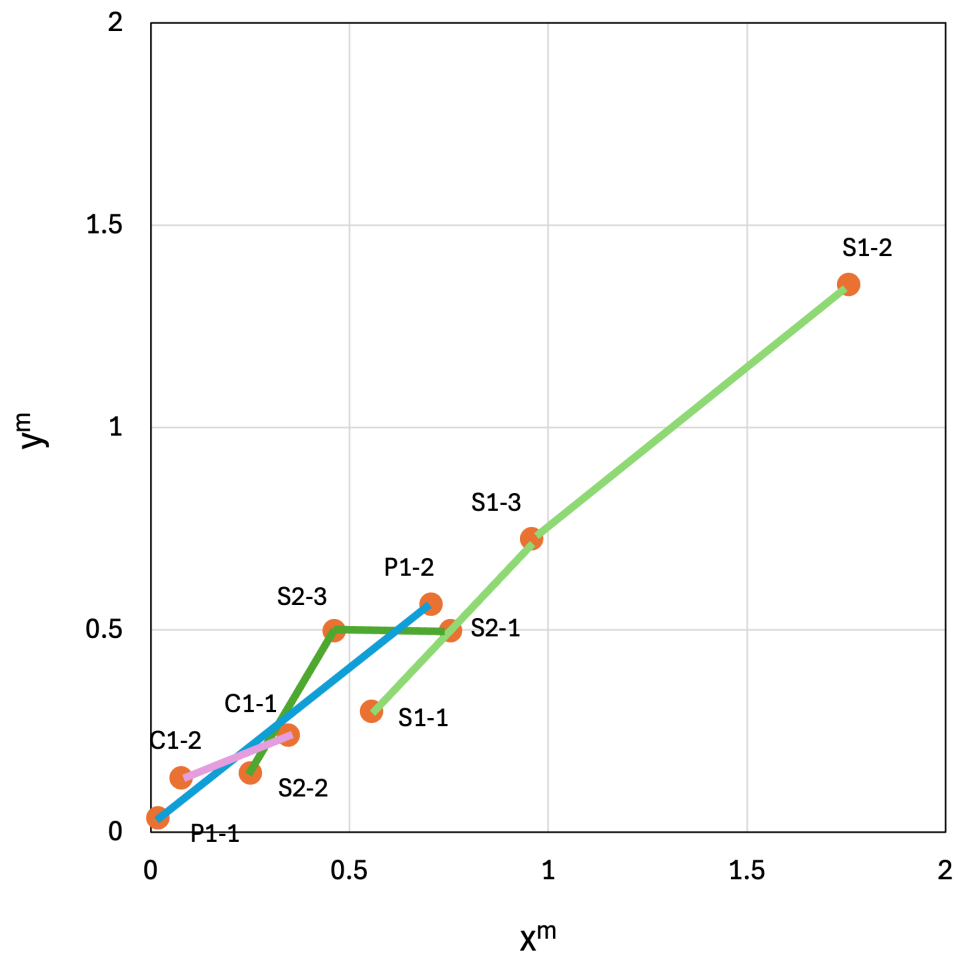

**Fig. 9.** Centroids of the ten individual anomalies in the modified Mahalanobis space.

## 5.4 Identification of anomaly types

Fig. 9 shows the centroids of the ten individual anomalies in the modified Mahalanobis space. All ten appear at distinct positions. The same anomaly type in machines of the same model (e.g., S1-2 and S2-2) also appears at different positions, showing that a single microphone can distinguish the machine in which an anomaly occurs. Several anomalies (P1-1, C1-2, S2-2, and S2-3), however, lie close together within $x^m$, $y^m \leq 0.4$, and P1-1 lies very close to the origin (0.019, 0.034). Moreover, for almost all anomalies the shift from the normal centroid points in the same direction ($\varphi_4$ decreasing, $\varphi_5$ increasing), so the two-dimensional map effectively distinguishes anomalies by the magnitude of the shift along this direction. The five pairs of simultaneous anomalies also appear at distinct positions (Fig. 10).

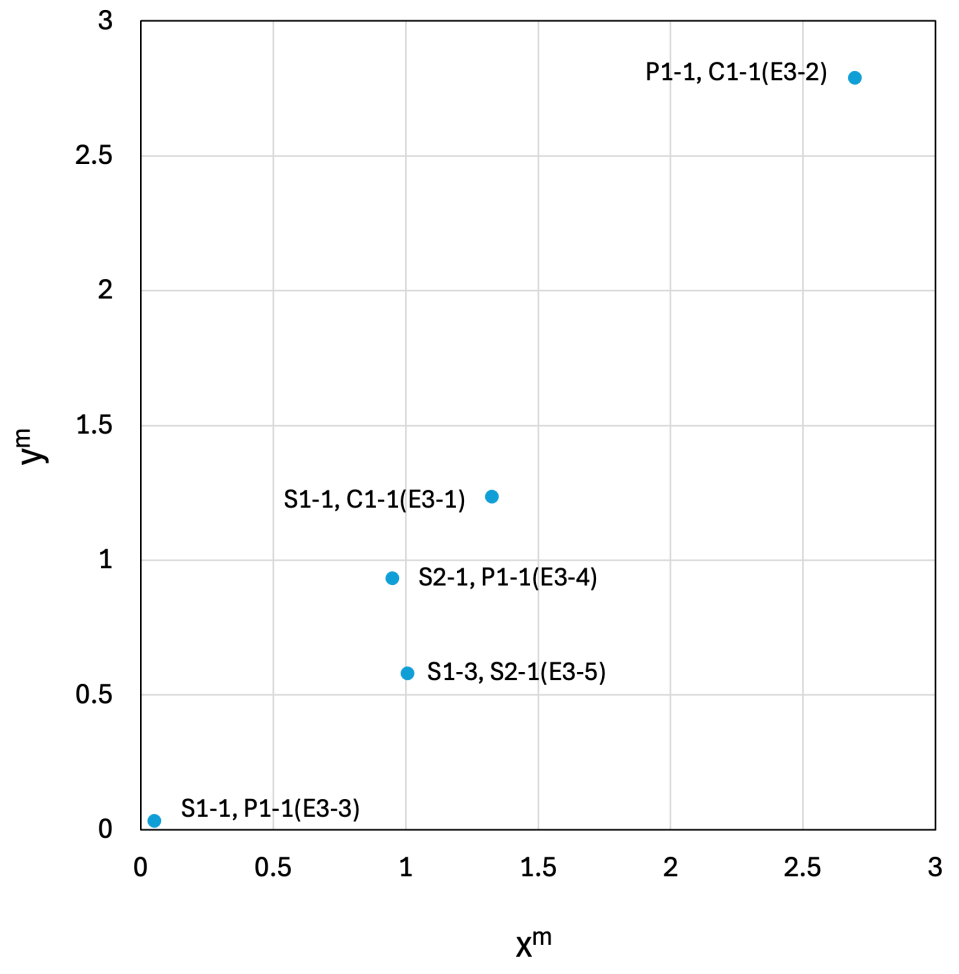

**Fig. 10.** Centroids of the five pairs of simultaneous anomalies in the modified Mahalanobis space.

### 5.5 Test of linear superposition

We tested whether the position of a simultaneous anomaly can be explained by the superposition of the corresponding individual anomalies. Because the coordinates of Eq. (4) take absolute values and are divided by class-specific $\sigma_k$, they cannot be tested for additivity directly. We therefore used the signed centroid shifts $d_k$ before division by $\sigma_k$. If the effects of anomalies a and b add linearly, the shift of the simultaneous anomaly should satisfy

$$d_{a+b} = d_a + d_b, \quad d_k = \mu_k - \mu_0 \tag{7}$$

Shifts of individual anomalies were computed relative to the normal centroid of Exp. 2, and those of simultaneous anomalies relative to the normal centroid of Exp. 3. Table 6 compares the observed shifts with the prediction of Eq. (7). For reference, the normal centroids of Exp. 2 and Exp. 3 differ by 0.17, which indicates the session-to-session variation.

**Table 6.** Centroid shifts ($\varphi_4$, $\varphi_5$) of simultaneous anomalies and the prediction of linear superposition.

| Pair | Observed $d_{a+b}$ | Predicted $d_a + d_b$ | Deviation (/SE) | Magnitude ratio (obs./pred.) |
|---|---|---|---|---|
| S1-1 + C1-1 | (−0.730, 0.682) | (−0.848, 0.670) | 0.12 (1.7) | 0.92 |
| P1-1 + C1-1 | (−1.385, 1.434) | (−0.417, 0.408) | 1.41 (20.3) | 3.42 |
| S2-1 + P1-1 | (−0.777, 0.763) | (−0.609, 0.366) | 0.43 (6.1) | 1.53 |
| S1-1 + P1-1 | (−0.044, −0.030) | (−0.458, 0.213) | 0.48 (6.7) | 0.10 |
| S1-3 + S2-1 | (−0.486, 0.280) | (−1.428, 1.018) | 1.20 (17.1) | 0.32 |

We compared each deviation with its standard error, obtained from the standard errors of the centroids (standard deviation/$\sqrt{n}$; values in parentheses in Table 6). Because the centroids were computed over 0.25-s sub-segments, n is the number of sub-segments per class (about 1,200). For S1-1 + C1-1, the deviation was only 1.7 SE, consistent with linear superposition. The other four pairs deviated by more than 6 SE. For P1-1 + C1-1, the shift was amplified to 3.4 times the prediction (20.3 SE); notably, P1-1 alone is barely distinguishable from normal operation (shift 0.03), yet together with C1-1 the shift increased greatly. The shift was also amplified for S2-1 + P1-1 (1.5 times; 6.1 SE). Conversely, it was attenuated to 0.32 times the prediction for S1-3 + S2-1 (17.1 SE) and almost vanished for S1-1 + P1-1 (0.10 times; 6.7 SE). Because the 20 sub-segments within a sample are correlated, the true standard errors are larger. Even under the most conservative per-sample estimate ($n \approx 60$), P1-1 + C1-1 (4.5 SE) and S1-3 + S2-1 (3.8 SE) deviate clearly. Thus, simultaneous anomalies do not lie at the simple superposition of individual anomaly positions for at least two pairs, and for four pairs at the sub-segment level.

P1-1 + C1-1 is the only pair that the RBF kernel also detected well ($F_2 = 0.9$), consistent with the interpretation that classes far from normal are easier for classical kernels. QK25, in contrast, detected S1-1 + P1-1 with $F_2 \approx 1.0$ even though its shift in the $\varphi_4$–$\varphi_5$ plane almost vanished, suggesting that the quantum kernel captures differences spread across all six features that do not appear in the centroids of these two coefficients.

## 6. Discussion

### 6.1 Advantages of quantum kernels

Table 7 contrasts the RBF kernel with quantum feature maps. The RBF kernel deforms the feature space isotropically on the basis of Euclidean distance, whereas a quantum feature map measures similarity by the angle between quantum states, and the phases of rotation gates together with CNOT-induced interference fold the space. The poor RBF results suggest that normal and anomalous samples overlap in Euclidean space and cannot be separated by isotropic deformation, consistent with the anomalous signals being 1.1–1.7 dB weaker than normal operation relative to background noise. The Isolation Forest, which partitions along feature axes only, shows a similar limitation. Circuits without CNOT gates can be simulated efficiently classically, so the difference is specific to entangling feature maps.

**Table 7.** Classical RBF kernel versus quantum feature maps.

| Aspect | Classical RBF | Quantum feature map |
|---|---|---|
| Deformation of feature space | Isotropic expansion of Euclidean space | Folding by phase and interference |
| Similarity | Euclidean distance | Angle between states |
| Geometry | Flat, isotropic | Curved, folded |
| Source of nonlinearity | Gaussian function | Phase and interference |
| Behavior with small data | Depends on $\gamma$ | Depends on phase-space structure |

QK23 and QK24 differ in entanglement structure (one-directional versus bidirectional), whereas QK24 and QK25 differ only in input scaling (a factor of 1 versus 2). The improvement from QK23 to QK24 can therefore be attributed to bidirectional entanglement, and that from QK24 to QK25 to the kernel bandwidth rather than to higher expressivity; the bandwidth of quantum kernels is known to strongly affect their generalization [18], [19]. In Exp. 3, where Section 5.5 shows that anomaly positions shift nonlinearly, QK25 retained the highest $F_2$, whereas the isotropic RBF kernel failed for most pairs. This suggests that a bidirectional entangling feature map with an appropriate bandwidth is advantageous for such nonlinear configurations, although kernel-matrix analysis and a systematic sweep of the input scaling are needed to confirm the mechanism.

## 6.2 Nonlinear interactions in simultaneous anomalies

At least two mechanisms may explain why simultaneous anomalies do not follow superposition. The first is physical interaction in the measurement environment: the five machines operate in the same room, so vibration from one anomaly may propagate through the floor or frames, sound may interfere through reflections and reverberation, and an anomaly may alter the operating state of other machines through shared load or power. P1-1 + C1-1 (3.4 times the prediction) and S2-1 + P1-1 (1.5 times) suggest mutual reinforcement, whereas S1-1 + P1-1 (0.10 times) and S1-3 + S2-1 (0.32 times) suggest cancellation or masking. S1-1 + C1-1, in contrast, was nearly additive (0.92 times), so the presence and direction of interaction depend on the pair.

The second is nonlinearity in feature extraction. AR coefficients are estimated from the autocovariance of the signal; the autocovariance of a sum of two signals includes their cross-covariance, and the estimate involves the inverse of the autocovariance matrix. Even if the acoustic signals added linearly, the AR coefficients, and hence the positions in the modified Mahalanobis space, would not. The two mechanisms can be separated by comparing recorded simultaneous anomalies with synthetic signals formed by adding the separately recorded individual anomalies and processing them through the same pipeline: if the synthetic signals also deviate from the superposition range, feature extraction is the main cause; if only the recorded signals deviate, acoustic interaction is. In either case, the ability of the quantum kernel to detect anomalies whose features are not simple sums of individual anomalies supports its advantage for simultaneous multi-anomaly detection.

### 6.3 Limitations

First, performance in Exp. 3 was lower than in Exp. 2, and not all scores reached 1.0. Second, several anomaly classes lie close together near the origin of the modified Mahalanobis space, and separating them requires further work. Third, only five pairs were evaluated, the standard errors of the centroids assume independent sub-segments, and the normal centroids of Exp. 2 and Exp. 3 differ by 0.17, so session-to-session variation must be taken into account. In addition, the identification map uses only two AR coefficients, and because the anomaly shifts point in nearly the same direction, separation near the origin is limited; identification using the Mahalanobis distance over all six coefficients is left for future work. Fourth, the quantum kernels were computed with a state-vector simulator, so the effect of hardware noise is not assessed. Fifth, the quantiles and threshold of the exclusion rule were set empirically, considering the limited data and the need to retain anomaly information; the number of excluded samples and the sensitivity of the results to these settings should be reported. Sixth, each condition consists of 60 consecutive samples cut from a single 5-min recording, so the samples are temporally correlated and cannot be regarded as independent trials, and differences between anomaly classes may include recording-specific conditions as well as the anomalies themselves. Indeed, the normal test samples showed false-alarm rates of 7–9% for the quantum kernels, much higher than expected from $\nu = 0.1$ under the 15-of-20 rule if sub-segments were independent, which suggests correlation between sub-segments and a difference between the separate training and test recordings of normal operation. Recordings made on different occasions are needed to evaluate session-to-session variation, and because the multinomial bootstrap assumes independent samples, the reported confidence intervals may be optimistic.

## 7. Conclusion

We investigated simultaneous multi-anomaly detection from a single microphone using AR features, quantum-kernel OCSVMs, and a modified Mahalanobis space, and compared three quantum feature maps with an RBF-kernel OCSVM and an Isolation Forest under the same two-stage decision rule. Performance improved from QK23 to QK24 with bidirectional entanglement and further with QK25, which is equivalent to QK24 with doubled input scaling (bandwidth). QK25 detected individual anomalies almost perfectly ($F_2 = 0.994$) and simultaneous two-anomaly pairs with $F_2 = 0.872$, outperforming both classical baselines in threshold-dependent and threshold-independent metrics. The modified Mahalanobis space separated both individual and simultaneous anomalies, but simultaneous anomalies did not lie at the superposition of individual positions, suggesting nonlinear acoustic interactions. Future work will improve the separation of spatially proximate anomaly clusters, characterize these interactions using synthetic superposition data, examine the kernel bandwidth systematically, and evaluate the method on quantum hardware toward practical deployment.

## Acknowledgments

This work was supported by the New Energy and Industrial Technology Development Organization (NEDO), Japan, under Grant No. JPNP23003.